\documentclass[twocolumn,trackchanges]{aastex631}
\usepackage{amsmath}
\usepackage{threeparttable}
\begin{document}

\title{CZ Aqr: an oscillating eclipsing Algol-type system composed of a $\delta$ Sct primary star and a subgiant star in a quadruple system}

\correspondingauthor{Wen-Ping Liao}
\email{$^{*}$liaowp@ynao.ac.cn}	
\author{Qi-Huan Zeng}
\affiliation{Yunnan Observatories, Chinese Academy of Sciences (CAS), Kunming, 650216, Yunnan, China}
\affiliation{University of Chinese Academy of Sciences, No.1 Yanqihu East Rd, Huairou District, 101408, Beijing, China}

\author{Wen-Ping Liao}
\affiliation{Yunnan Observatories, Chinese Academy of Sciences (CAS), Kunming, 650216, Yunnan, China}
\affiliation{University of Chinese Academy of Sciences, No.1 Yanqihu East Rd, Huairou District, 101408, Beijing, China}

\author{Sheng-Bang Qian}
\affiliation{Department of Astronomy, School of Physics and Astronomy, Yunnan University, 650091 Kunming, China}
\affiliation{Key Laboratory of Astroparticle Physics of Yunnan Province, Yunnan University, 650091 Kunming, China}

\author{Lin-Jia Li}
\affiliation{Yunnan Observatories, Chinese Academy of Sciences (CAS), Kunming, 650216, Yunnan, China}

\author{Ping Li}
\affiliation{Yunnan Observatories, Chinese Academy of Sciences (CAS), Kunming, 650216, Yunnan, China}
\affiliation{University of Chinese Academy of Sciences, No.1 Yanqihu East Rd, Huairou District, 101408, Beijing, China}

\author{Zhao-Long Deng}
\affiliation{Yunnan Observatories, Chinese Academy of Sciences (CAS), Kunming, 650216, Yunnan, China}
\affiliation{University of Chinese Academy of Sciences, No.1 Yanqihu East Rd, Huairou District, 101408, Beijing, China}



\begin{abstract}
Eclipsing Algol-type systems containing a $\delta$ Scuti (hereafter $\delta$ Sct) star enable precise determination of physical parameters and the investigation of stellar internal structure and evolution. We present the absolute parameters of CZ  Aquarius (hereafter CZ Aqr) based on TESS data. CZ Aqr has an orbital period of 0.86275209 d, a mass ratio of 0.489 (6), and the secondary component nearly fills its Roche lobe. $O-C$ analysis reveals a downward parabolic trend and a cyclical variation with a period of 88.2 yr. The downward parabola suggests a long-term decrease in the orbital period with $\dot{P}$ = -3.09$\times$$10^{-8}$ d $\textrm{yr}^{-1}$. The mass loss rate is estimated to be 4.54$\times$$10^{-9}$ M$_{\odot}$ $\textrm{yr}^{-1}$, which possibly due to magnetic stellar wind or hot spot. The cyclical variation might be caused by the light travel time effect via the presence of a third body with a minimum mass of $M_{3min}$ = 0.312 (21) M$_{\odot}$. Additionally, there are two possible celestial bodies in a 2:7 resonance orbit around CZ Aqr. The asymmetric light curve is explained by adding a hot spot on the surface of the primary star. After removing the binary model, 26 frequencies were extracted from TESS data. Two radial modes were newly identified among three possible independent frequencies. Our results show that the eclipsing Algol-type system is composed of a $\delta$ Sct primary star and a subgiant star in a quadruple system.
\end{abstract}



\section{Introduction} \label{sec:intro}

Eclipsing binary stars are particularly reliable for the determination of stellar physical parameters. The presence of pulsating stars within these systems offers a valuable opportunity to employ binary analysis techniques in the study of variable stars. Approximately 70$\%$ of stars in the universe are members of binary or multiple star systems
\citep{2009AJ....137.3358M,2011IAUS..272..474S}. Eclipsing Algol-type systems with a $\delta$ Sct component have been characterized as oscillating eclipsing Algol-type (oEA) systems  
\citep{2004A&A...419.1015M}. The primary component of the oEA system is typically an A-F type main sequence (MS) star, while the secondary component is a giant or subgiant star that is filling, or nearly filling, its Roche lobe. The Algol-type system can undergo rapid mass transfer from a more massive star to another, resulting in a reversal of the mass ratio and the previously more massive component becoming the less massive component
\citep{1998A&AT...15..357P,2017MNRAS.470..915K}. In the mass-transferring binary system, the pulsating primary star can serve as compelling evidence of binary interaction, facilitating further exploration into the structure and evolution of stars
\citep{2020ApJ...895..136C}. When a pulsating variable star is part of an eclipsing binary system, critical parameters (e.g., mass, radius, luminosity) of the pulsating component can be derived from the system's eclipsing characteristics.

The $\delta$ Sct-type star lies at the intersection of the instability strip (IS) of classical Cepheids and the MS band in the Hertzsprung-Russell (H-R) diagram. They are either in the central hydrogen-burning phase of the MS or have just left the MS, with spectral types A-F, a mass range of 1.5 to 2.5 M$_{\odot}$, a pulsating period range of 0.02-0.3 $d$, and a surface effective temperature range of 6900 to 8900 K
\citep{2017ampm.book.....B}. Typically, they can exhibit radial and non-radial p-mode oscillations in the stellar envelope due to pressure variations, and g-mode oscillations in the stellar interior due to gravity variations
\citep{2010aste.book.....A,2018MNRAS.476.3169B}. The interaction of g and p mode characteristics occurs via a succession of avoided crossings, which facilitate the interaction between the modes
\citep{1975PASJ...27..237O}. Mixed modes are commonly detected in stars that have evolved off the main sequence and are undergoing hydrogen shell burning, primarily due to the substantial density gradient exterior to the core, which couples the g-mode and p-mode pulsation cavities
\citep{2010aste.book.....A}.
\cite{2022ApJS..263...34C} discovered 123 new $\delta$ Sct-type stars in eclipsing binaries based on Transiting Exoplanet Survey Satellite (TESS) data. With the release of TESS's massive data, more and more candidates are expected to be identified for further study. The majority of the binary members-$\delta$ Sct stars commonly lie within the MS band and approach the zero-age main sequence (ZAMS) boundary, suggesting that their inherent instabilities emerge at earlier stages of their evolution compared to solitary stars
\citep{2012MNRAS.422.1250L}. The intricate pulsational characteristics and evolutionary progression of $\delta$ Sct-type stars make them a powerful tool for investigating stellar evolution.

The light travel time effect (LTTE) is a common phenomenon when a close-in third body orbits a central eclipsing binary, causing the binary to orbit around the barycenter of the triple system. The arrival times of the eclipses should change cyclically as the binary periodically approaches or moves away from the observer
\citep{2013AJ....146...28Z}. These additional objects may promote the formation of binaries by acquiring or absorbing angular momentum
\citep{2022ApJ...927..183L}.
Therefore, it is important to investigate these physical phenomena. In recent years, the launches of the TESS and the Kepler Space Telescope have provided a wealth of continuous and high-precision data to investigate whether LTTE is also commonly present in oEA systems.

CZ Aquarius (CZ Aqr) was identified as a short-period Algol-type eclipsing binary in the southern sky with an A5-type primary star, which exhibit a long-term increase in the orbital period as revealed by O-C analysis
\citep{1986IBVS.2865....1B}. \cite{2012MNRAS.422.1250L} identified a potential third body through O-C analysis and demonstrated that it was an oscillating eclipsing Algol-type binary with a principal pulsation frequency of 35.5 $\textrm{d}^{-1}$. He also determined its absolute parameters through photometric analysis. In an eclipsing Algol-type system, when the secondary component fills or overflows its Roche lobe, mass is transferred to the primary component, and the orbital period typically either increases or remains constant
\citep{1955AnAp...18..379K}. In this paper, we combined TESS light curves with eclipsing times from the literature to analyze the orbital period variation and the pulsational properties of CZ Aqr. Additionally, the absolute parameters of the binary were calculated.

\section{INVESTIGATION OF THE ORBITAL PERIOD VARIATION FOR CZ AQR} \label{OC}

\begin{figure}
	\centering
	\includegraphics[width=0.5\textwidth, angle=0]{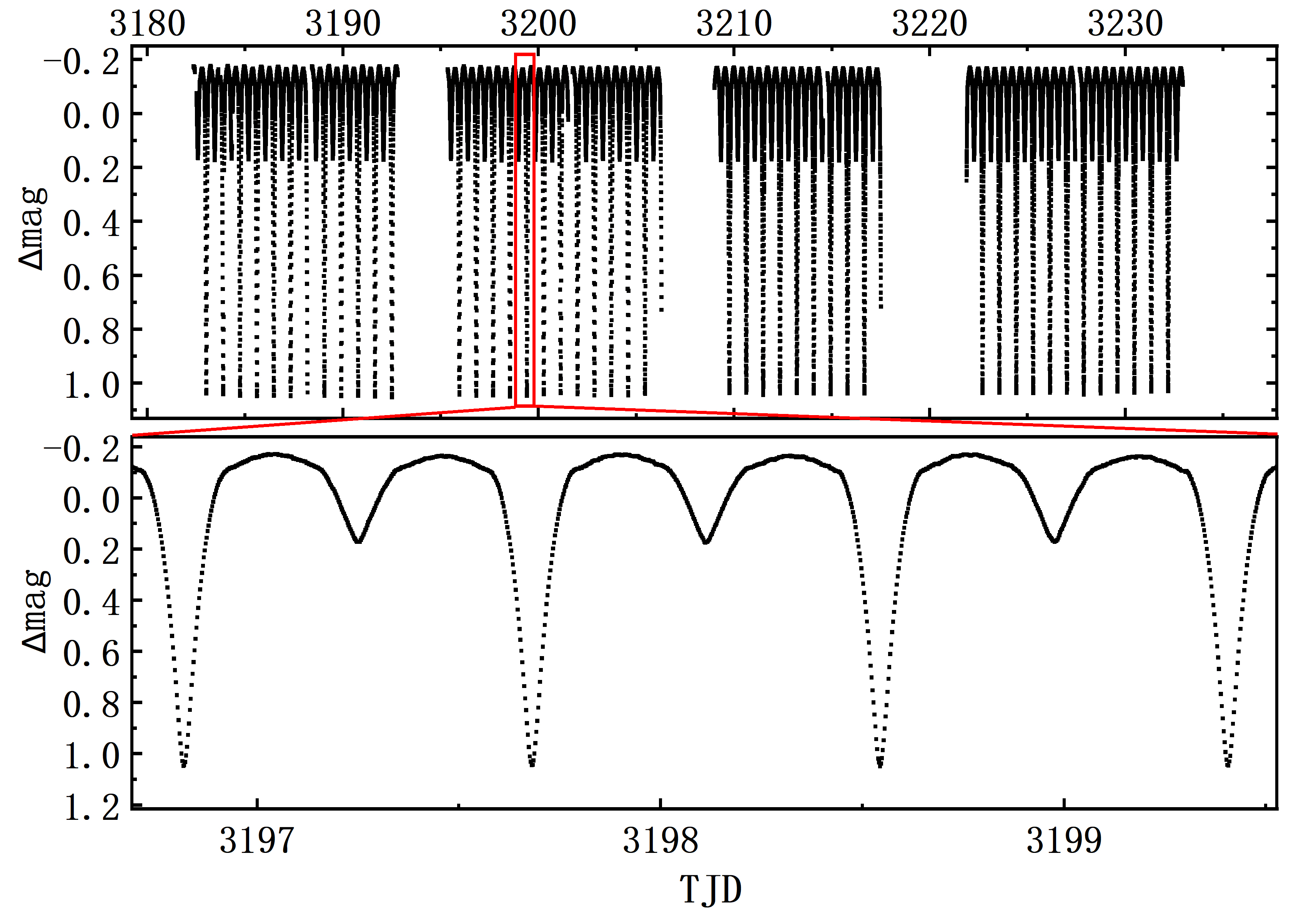}	
	\caption{The processed fragment of the TESS light curve for CZ Aqr (TJD = BJD - 2457000).}
	\label{fig:figure1}%
\end{figure}

TESS is equipped with four identical cameras, and the northern and southern ecliptic hemispheres are divided into 13 partially overlapping sectors. Each sector is observed continuously for 27.4 days
 \citep{2015JATIS...1a4003R}. The light curves of CZ Aqr were observed by TESS across five sectors (02, 29, 42, 69, and 70) from TJD 1354 to 3232 (where TJD = BJD-2457000). This dataset consists of roughly 53,000 data points and covers five years. Sectors 02, 42, and 70 have a short cadence of 120 s, while sector 69 has an exposure time of 200 s, and sector 29 has an exposure time of 600 s. The data for Simple Aperture Photometry (SAP) are acquired from the Mikulski Archive for Space Telescopes (MAST) database. The SAP flux is preferred over the Pre-search Data Conditioned Simple Aperture Photometry (PDCSAP) flux because it is not subject to any pre-processing. The light curve data underwent three stages of processing. Initially, the flux was transformed into magnitude. Subsequently, the mean magnitude of each sector was subtracted to normalize the data. Finally, the light curve was converted into a phase curve. The processed fragment of the TESS light curve for CZ Aqr is shown in Fig. \ref{fig:figure1}.

\begin{figure*}
	\centering
	\includegraphics[width=0.9\textwidth, angle=0]{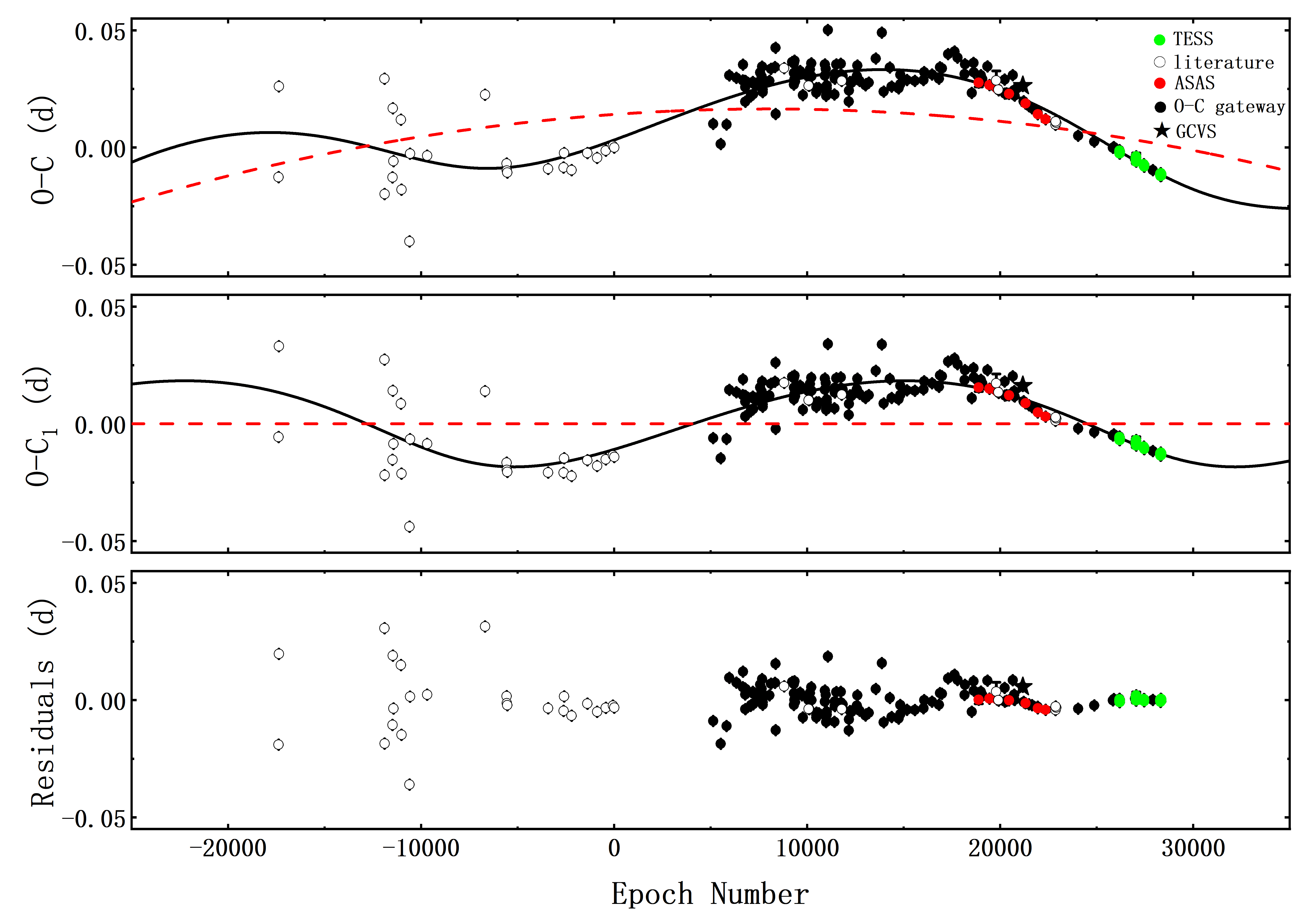}	
	\caption{The $O-C$ curve is fitted with an eccentric orbit. The fitted results, including a downward parabola and cyclical variation, are shown in the upper panel; the cyclical variation alone is depicted in the middle panel; and the residuals are displayed in the bottom panel.}
	\label{fig:figure2}%
\end{figure*}

Typically, changes in the orbital period of close binaries are induced by the transfer or loss of mass and angular momentum from the binary system, LTTE, and magnetic activity. The eclipse timing variation (ETV) method (also called as ‘$O-C$ analysis’)
\citep{2022A&A...663A.137L} was used to analyze the orbital period variation of CZ Aqr. To obtain additional light minima, we also collected light minima from the $O-C$ gateway\footnote{http://var2.astro.cz/ocgate/}. We converted data spanning many cycles into a phase and subsequently used parabola fitting to get the light minima for TESS
\citep{2021MNRAS.505.6166S,2022ApJ...924...30L}. Finally, we obtained 113 light minima from TESS, and additionally collected 7 light minima from ASAS, 1 light minimum from GCVS, and 169 light minima from the $O-C$ gateway from
\cite{2003IBVS.5438....1D},
\cite{2006IBVS.5676....1K},
\cite{2010IBVS.5958....1L},
\cite{1986IBVS.2865....1B}, and
\cite{2011IBVS.5960....1D}. All times of light minima are listed in Table \ref{tab:tab 6}. The $O-C$ (where ‘$O$’ denotes the observed times of the eclipse egress and ‘$C$’ denotes the times computed using a linear ephemeris) values are calculated using the linear ephemeris formula:
\begin{eqnarray}\label{equation(1)}
MinI (HJD) =2435778.3570 + 0^{\textrm{d}}.86275209\times{E}
\end{eqnarray}

where $HJD_0$ was obtained from 
\cite{1986IBVS.2865....1B} and the value of the orbital period was obtained from the International Variable Star Index(VSX)\footnote{VSX:https://www.aavso.org/vsx/}. The calculation of the light minima resulted in the O-C diagram shown in Fig. \ref{fig:figure2}. In the $O-C$ curve, the green dots represent data from TESS, the red dots represent data from ASAS, the black pentagram represents data from GCVS, and the open circles represent data from the literature. Additionally, the dark dots represent data from the $O-C$ gateway. In the $O-C$ curve, a clear periodic oscillation is evident, which may be attributed to the LTTE of a third body
\citep{2010MNRAS.405.1930L,2010Ap&SS.329..113Q,2024MNRAS.tmp.1826J,2016AJ....152...26M}. Therefore, we utilize the equation presented by 
\cite{1952ApJ...116..211I} along with the Kepler equation to fit the $O-C$ diagram:
\begin{eqnarray}\label{equation(2)}
\begin{aligned}
\mathrm{O}-\mathrm{C}(\mathrm{d})=\Delta \mathrm{T}_{0}+\Delta \mathrm{P}_{0} \times \mathrm{E}+\frac{\beta}{2} \mathrm{E}^{2}+\tau,
\end{aligned}
\end{eqnarray}

\begin{eqnarray}\label{equation(3)}
\begin{aligned}
\tau=\mathrm{A}\left[\left(1-e_{3}^{2}\right) \frac{\sin (v+\omega)}{1+\mathrm{e}_{3} \cos v}+\mathrm{e} \sin \omega\right]   
\end{aligned}
\end{eqnarray}

\begin{eqnarray}\label{equation(4)}
\begin{aligned}
\tau=\mathrm{A}\left[\sqrt{1-e_{3}^{2}} \sin \mathrm{E}^{*} \cos \omega+\cos \mathrm{E}^{*} \sin \omega\right],   
\end{aligned}
\end{eqnarray}

\begin{eqnarray}\label{equation(5)}
\mathrm{M}=\mathrm{E}^{*}-\mathrm{e}_{3} \sin \mathrm{E}^{*}=\frac{2 \pi}{\mathrm{P}_{3}}\left(\mathrm{t}-T_{3}\right)
\end{eqnarray}

where $\Delta$$T_{0}$ and $\Delta$$P_{0}$ are the correction values for the initial epoch and orbital period of the binary, $\beta$ is the long-term change in the orbital period (d $\textrm{cycle}^{-1}$), and $\tau$ represents the cyclical variation of the component due to the LTTE. Here, $A$ = $a_{12}$sin $i_{3}$/c, where $a_{12}$sin $i_{3}$ is the projected semimajor axis and c is the speed of light. $e_{3}$ is the eccentricity of a supposed third body, $\nu$ is the true anomaly of the position of the eclipsing pair's mass centre on the orbit, $\omega$ is the longitude of the periastron of the eclipsing pair's orbit around the third body. Additionally, $M$, $E^{*}$ and $P_{3}$ are the mean anomaly, the eccentric anomaly, and the period of a supposed third body, respectively, and $T_{3}$ is the time of periastron passage
\citep{2010PASJ...62.1109L,2021AJ....161..193L}. All of these parameters are obtained by fitting the $O-C$ curve using equations (\ref{equation(2)}) through (\ref{equation(5)}) and are listed in Table \ref{tab:tab 1}. The fitted $O-C$ curve and residuals are shown in Fig. \ref{fig:figure2}, which shows a downward parabolic trend and cyclical variation. We obtain $A$ = 0.0185 (±0.0016) day, $P_{3}$ = 88.2 (±5.2) yr and $\beta$ = -7.3 (±1.5)$\times$$10^{-11}$ d $\textrm{cycle}^{-1}$ = -3.09 (±0.63)$\times$$10^{-8}$ d $\textrm{yr}^{-1}$.

\begin{figure}
	\centering
	\includegraphics[width=0.55\textwidth, angle=0]{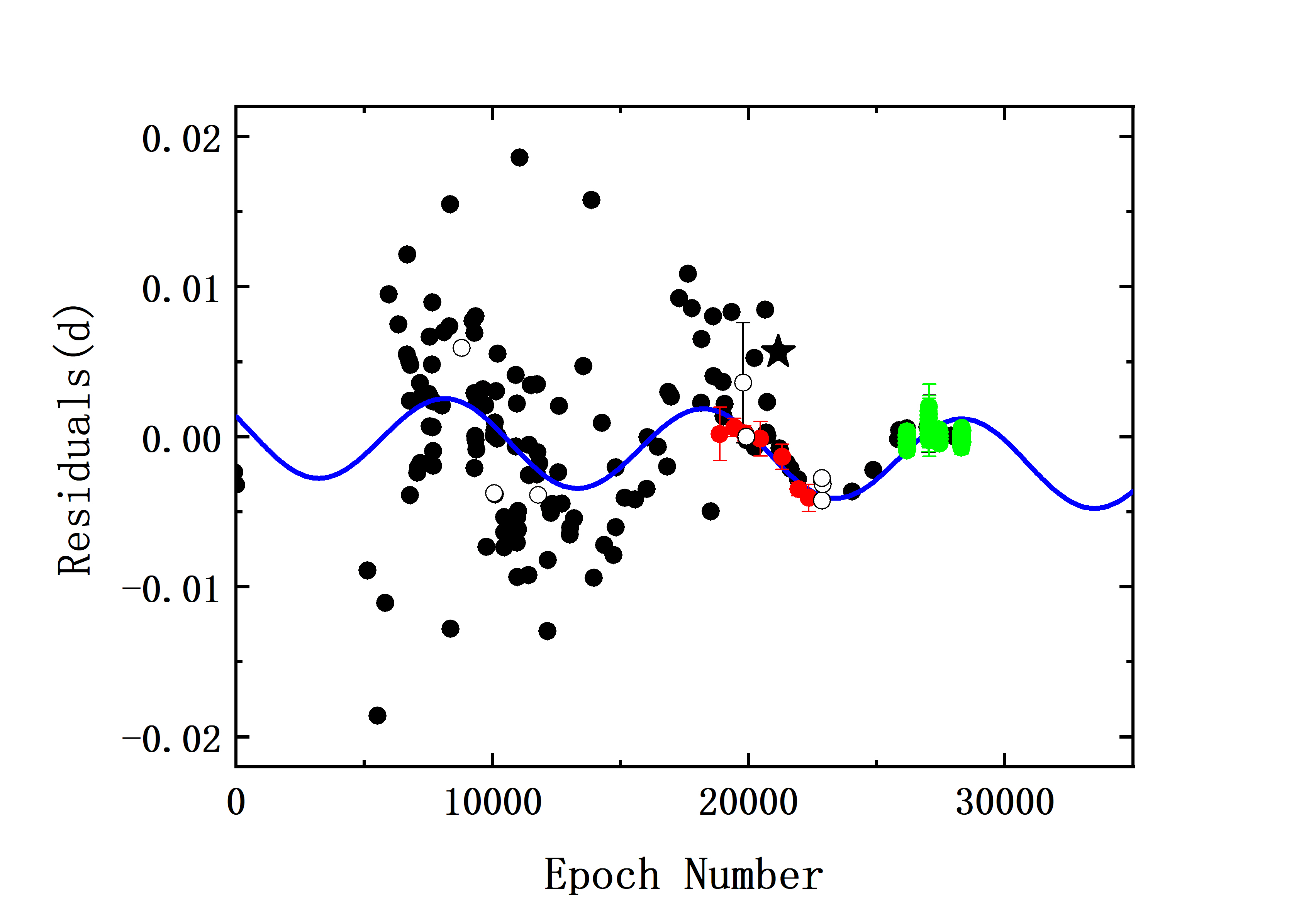}	
	\caption{The residuals diagram, containing more than 5,000 cycles, displays the fitting results of the cyclical variation.}
	\label{fig:figure3}%
\end{figure}

In the $O-C$ analysis, we also discerned for the first time that another potential periodic oscillation may exist when analyzing more than 5,000 epochs, as shown in Fig. \ref{fig:figure3}. We attempted to fit the residuals using a cyclical variation. The period of the cyclical variation is 10,090 cycles (approximately 23.85 yr), suggesting the possible presence of a fourth body. Therefore, the orbital period ratio is 0.27 (approximately 2:7), indicating that two additional celestial bodies might orbit CZ Aqr binary pair in a 2:7 resonance orbit
\citep{2019RAA....19..107W,2016RAA....16...94L}. 

\begin{figure}
	\centering
	\includegraphics[width=0.55\textwidth, angle=0]{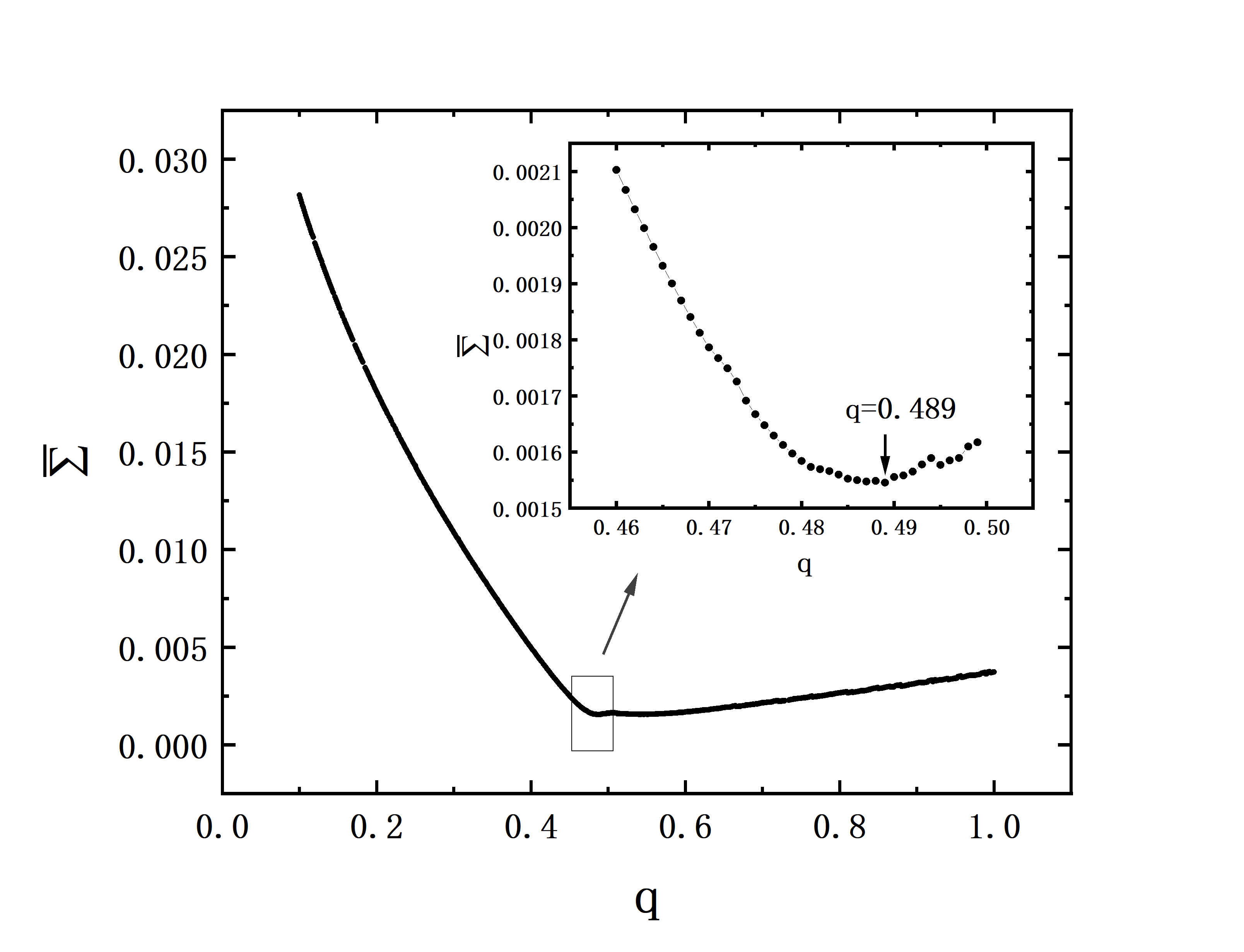}	
	\caption{The relationship between the mean residuals and the mass ratio $q$ indicates a minimum value near $q$ = 0.489 for CZ Aqr (Model 5).}
	\label{fig:figure4}%
\end{figure}

\begin{figure}
	\centering
	\includegraphics[width=0.55\textwidth, angle=0]{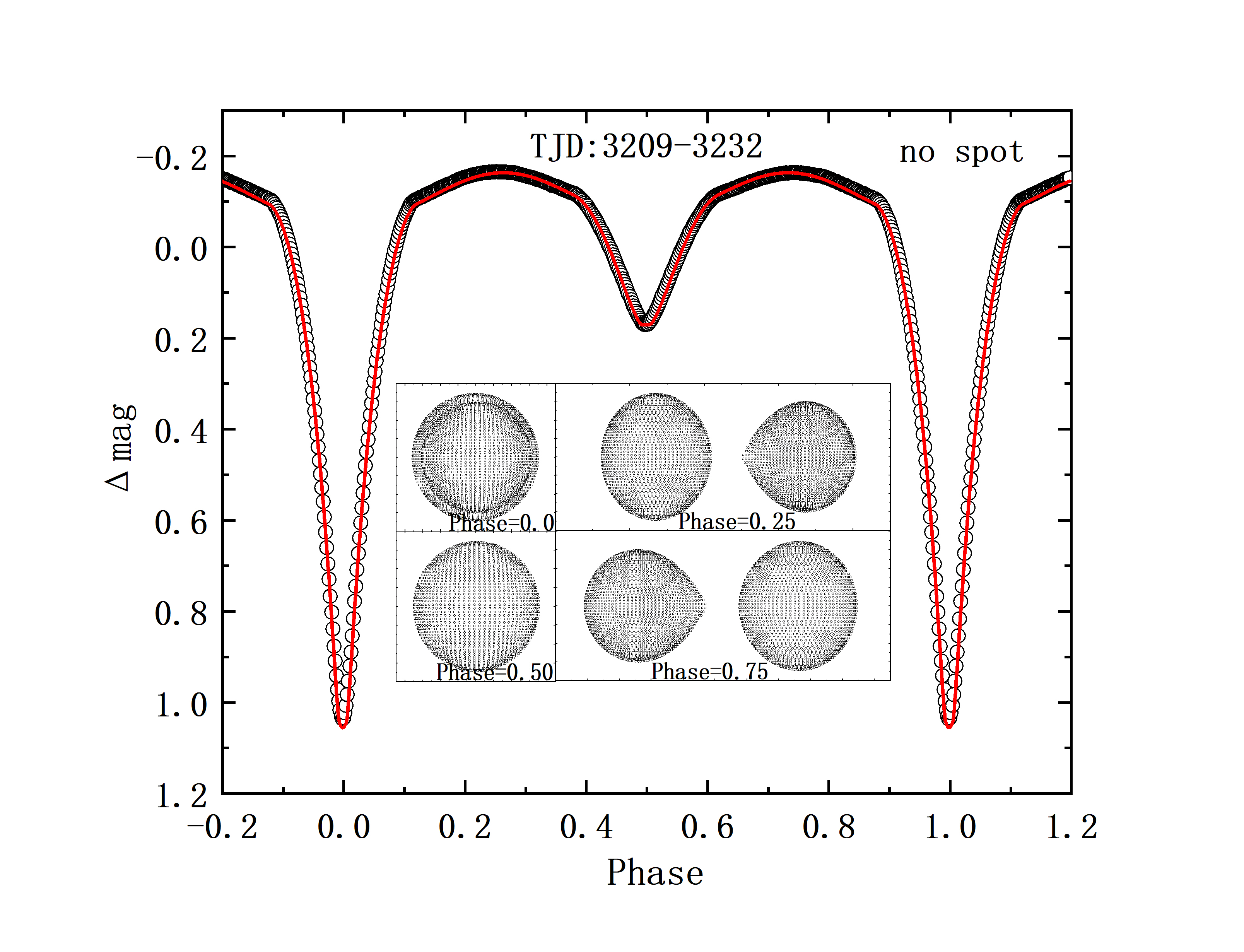}	
	\caption{The symmetric light curve for CZ Aqr is presented. The black open circles represent the average light curve, the red solid line denotes the theoretical light curve, and the geometrical structure diagrams are displayed in the center.}
	\label{fig:figure5}%
\end{figure}

\section{PHOTOMETRIC SOLUTIONS} \label{WD}

The Wilson-Devinney (W-D) program
\citep{1971ApJ...166..605W,1990ApJ...356..613W,2007ApJ...661.1129V,2014ApJ...780..151W} is employed to analyze the light curves. As shown in the bottom panel of Fig. \ref{fig:figure1}, the two peaks of the TESS light curve are at the same level around TJD 3198 (MAX I = MAX II). Consequently, the data from sector 70 is selected for analyzing the basic solutions without any spots. Taking into account the impact of pulsation, a mean light curve is generated by averaging 13,700 data points into 500 data points, yielding 0.002 phase per data point.

The primary spectral type of CZ Aqr was previously determined to be A5
\citep{1986IBVS.2865....1B,2012MNRAS.422.1250L,2010arXiv1002.2729Z}. The magnitudes of CZ Aqr in the B and V bands are found to be 11.1 mag and 10.98 mag
\citep{2010arXiv1002.2729Z}, thus a B-V value of 0.12 mag could be computed. Using the relationship between B-V color and effective temperature in Table 3 given by
\cite{1996ApJ...469..355F}, we estimate that the surface temperature of the primary component to be 8200 K. Both this estimated temperature and that given by
\cite{2012MNRAS.422.1250L}, are consistent with the classification of an A5-type star based on the relations provided by
\cite{2000asqu.book.....C}. The surface temperature of the primary component is fixed in the binary modeling. Given that the primary star is an early-type star, its bolometric albedo could be set as $A_{1}$ = 1.00 and $A_{2}$ = 0.50
\citep{1969AcA....19..245R}, along with gravity-darkening coefficients $g_{1}$= 1.00 and $g_{2}$= 0.32 
\citep{1967ZA.....65...89L}. Given the configuration of oEA, we attempted to fit the light curve using model 2 (detached binary) and model 5 (semidetached binary with star 2 filling its Roche lobe). However, model 2 did not converge. We employed the q-search method to derive the initial mass ratio. For model 5, the convergent solutions were obtained when the mass ratio was fixed at values ranging from 0 to 1 with a step of 0.001, and the minimal mean residuals $\bar{\Sigma}$ were achieved at a $q$ value of 0.489 (±0.006). The q-search diagram is presented in Fig. \ref{fig:figure4}. Then, we set q as an adjustable parameter and run the W-D program for model 5. The other adjustable parameters include the orbital inclination $i$, the mass ratio $q$ ($M_{2}$/$M_{1}$), the luminosity $L_{1}$ of star 1, the temperature $T_{2}$ of star 2, and the dimensionless potential $\Omega_{1}$ of star 1. We also attempted to fit the light curve by including a third light but found no convergent solutions. The photometric solutions of CZ Aqr are listed in Table \ref{tab:tab 2}, where the errors of $L_{1}$/($L_{1}$+$L_{2}$), $f_{1, \,2}$, and $\rho_{1,\,2}$ are calculated using error propagation, while the errors for other adjustable parameters are provided by the W-D program. The volume filling factors (= $V_{star}$ / $V_{L}$) of the primary and secondary components are 50.97$\%$ and 99.84$\%$, respectively. It is suggested that CZ Aqr could be a semidetached binary with its secondary component filling its Roche lobe. The photometric analysis results from both
\cite{2012MNRAS.422.1250L} and this paper do not detect a third light. This implies that the third body has a negligible light contribution, which cannot be detected. The fitting results of the symmetric light curve are shown in Fig. \ref{fig:figure5}, which also displays the geometrical structure diagrams.

\begin{table*}
\begin{center}
\footnotesize
\caption{Orbital parameters of the third body for CZ Aqr. \label{tab:tab 1}}
\setlength{\tabcolsep}{0.68cm}{
       \begin{tabular*}{0.75\linewidth}{@{}ccc@{}}
\hline       
Parameters                                & Value                   & Unit        \\
\hline
$T_{0}$ (Revised epoch)                         & 2435778.3711 ($\pm$0.0020)   & HJD         \\
$P_{0}$ (Revised period)                        & 0.86275267 ($\pm$0.00000021) & days        \\
$\beta$ (Long-term change of the orbital period) & -7.3 ($\pm$1.5)$\times$$10^{-11}$   & d\,cycle$^{-1}$            \\
$A$ (Amplitude)                              & 0.0185 ($\pm$0.0016)       & d \\
$e_{3}$ (Eccentricity)                          & 0.20 ($\pm$0.12)       & -           \\
$P_{3}$ (Orbital period)                        & 88.2 ($\pm$5.2)             & yr          \\
$a_{12}$$\sin i_{3}$ (Projected semi-major axis)       & 27.5 ($\pm$1.1)              & au          \\
$f$(m) (Mass function)                       & 0.0042 ($\pm$0.0012)         & M$_{\odot}$         \\
$M_{3min}$ (Mass)                               & 0.312 ($\pm$0.021)           & M$_{\odot}$         \\
\hline
\end{tabular*}}
\end{center}
\end{table*}

\begin{table}
\caption{The photometric solutions of CZ Aqr (Model 5) are obtained by using the W-D program. Most parameters are dimensionless, except for those previously specified.\label{tab:tab 2}}
\setlength{\tabcolsep}{1.4cm}{
       \begin{tabular*}{\linewidth}{@{}cc@{}}
\hline
Parameters             & Model\,5       \\
\hline
$g_{1}$                & 1.00        \\
$g_{2}$               & 0.32        \\
$A_{1}$               & 1.00        \\
$A_{1}$               & 0.50        \\
q ($M_{2}$/$M_{1}$)     & 0.4866(13)  \\
$T_{1}$ (K)             & 8200   \\
$T_{2}$/$T_{1}$      & 0.67927(60) \\
$i$($^{\circ}$)        & 89.79(15)  \\
$\Omega_{1}$        & 3.3818(31)  \\
$\Omega_{2}$          & 2.85   \\
$L_{1}$/($L_{1}$+$L_{2}$)  & 0.79920(16) \\
$r_{1}$(pole)    & 0.34231(37) \\
$r_{1}$(side)    & 0.35334(42) \\
$r_{1}$(back)   & 0.36369(49) \\
$r_{2}$(pole)     & 0.29765(22) \\
$r_{2}$(side)   & 0.31062(23) \\
$r_{2}$(back)   & 0.34314(23) \\
$f_{1}$$^{a}$          & 0.5097(13)  \\
$f_{2}$$^{a}$         & 0.9984(26)  \\
$\rho_{1}$($\rho_{\odot}$)$^{b}$          & 0.27465(43)  \\
$\rho_{2}$($\rho_{\odot}$)$^{b}$         & 0.18271(11)  \\
$\bar{\Sigma}$ (mean residuals)            & 0.0014    \\
\hline
\end{tabular*}
\begin{tablenotes} 
		\item $^{a}$The ratio of star volume to Roche lobe volume ( = $V_{star}$ / $V_{L}$). 
  \item $^{b}$The mean density of star.
     \end{tablenotes}
}
\end{table} 

To obtain the absolute parameters of the system, we adopt the methods of
\cite{2024MNRAS.529.3113W} and
\cite{2022RAA....22c5024X}. The parallax $\varpi$ = 1.8343 ($\pm$0.0185) mas is provided by Gaia DR3, so the distance is approximately 545 ($\pm$5) pc. We take the brightest magnitude of CZ Aqr as the apparent magnitude in the V band ($m_{V}$ = 10.577 (±0.020) mag from ASAS-SN), the interstellar extinction $A_{v}$ = 0.075 mag from the NASA/IPAC Extragalactic 
Database, and the bolometric correction $BC_{v}$ $\approx$ -0.045 mag
\citep{2022AcA....72..195B}. Its absolute bolometric magnitude $M_{bol}$ is calculated using the following equation:

\begin{eqnarray}\label{equation(8)}
(m-M)_{v}=10-5 \lg \varpi+A_{v}, 
 \end{eqnarray}
\begin{eqnarray}\label{equation(9)}
M_{b o l}=M_{v}+B C_{v}
 \end{eqnarray}

we can then obtain the total luminosity of the binary $L_{T}$ using:

\begin{eqnarray}\label{equation(10)}
 L_{T}=L_{1}+L_{2}=(L/L_{\odot})=10^{0.4\times(4.74-\mathit{M}_{\rm bol})},
\end{eqnarray}
\begin{eqnarray}\label{equation(11)}
L_{T}/L_{\odot}=(ar_{1})^{2}\times(\frac{T_{1}}{T_{\odot}})^{4}+(ar_{2})^{2}\times(\frac{T_{2}}{T_{\odot}})^{4}
\end{eqnarray}

equation (\ref{equation(11)}) represents Stefan-Boltzmann’s law, where $T_{\odot}$ = 5772 K is the effective surface temperature of the Sun, and $r_{1}$ and $r_{2}$ are the relative radii of star 1 and star 2, respectively. The radius $r_{i}$ is calculated as ($r_{iploe}$$\cdot$$r_{iside}$$\cdot$$r_{iback}$)$^{1/3}$, where $r_{iploe}$, $r_{iside}$, and $r_{iback}$ are derived from the photometric solutions.

\begin{eqnarray}\label{equation(12)}
\frac{a^{3}}{P^{2}}=74.5(M_{1}+M_{2}),
\end{eqnarray}

\begin{eqnarray}\label{equation(13)}
q=\frac{M_{2}}{M_{1}}
\end{eqnarray}

combining Kepler's third law with the mass ratio $q$, we determined the masses of the binary components as $M_{1}$ = 1.60 (±0.06) M$_{\odot}$ and $M_{2}$ = 0.78 (±0.03) M$_{\odot}$, and the semi-major axis as $a$ = 5.09 (±0.07) R$_{\odot}$. The estimated absolute parameters are listed in Table \ref{tab:tab 3}, with the errors determined through error propagation.

\begin{table*}
\caption{The estimated absolute parameters of CZ Aqr. \label{tab:tab 3}}
\setlength{\tabcolsep}{1.4cm}{
       \begin{tabular*}{\linewidth}{@{}cccc@{}}
\hline
Parameters      & Value          & Parameters & Value        \\
\hline
$\varpi$ (mas)          & 1.8343 ($\pm$0.0185) & $d$ (pc)      & 545 ($\pm$5)       \\
$m_{V}$ (mag)              & 10.577 ($\pm$0.020)        & $M_{v}$ (mag)         & 1.819 ($\pm$0.030) \\
$M_{bol}$ (mag)            & 1.774 ($\pm$0.030)   & $BC_{v}$ (mag)        & -0.045        \\
$M_{1}$ (M$_{\odot}$)              & 1.60 ($\pm$0.06)     & $M_{2}$ (M$_{\odot}$)         & 0.78 ($\pm$0.03)  \\
$R_{1}$ (R$_{\odot}$)              & 1.80 ($\pm$0.03)   & $R_{2}$ (R$_{\odot}$)         & 1.61 ($\pm$0.02) \\
$L_{1}$ (L$_{\odot}$)              & 13.20 ($\pm$0.40)    & $L_{2}$ (L$_{\odot}$)         & 2.25 ($\pm$0.06)   \\
semi-major axis $a$ (R$_{\odot}$) & 5.09 ($\pm$0.07)     &            &              \\
\hline
\end{tabular*}}
\end{table*}

\section{FREQUENCY ANALYSIS} \label{PERIOD04}

We selected the short cadence (120 s) data from sector 70 to analyze the pulsational characteristics of CZ Aqr. We subtracted the binary's eclipsing light curve from the original one to obtain the pulsating light curve. The PERIOD04 software
\citep{2005CoAst.146...53L}, based on classical Fourier analysis, was employed to analyze the multiple frequencies. The Nyquist frequency and the frequency resolutions are $f_{Nyq}$= 359.387 $\textrm{d}^{-1}$ and $\delta$$f$ = 1.5/$\Delta$$T$ $\approx$ 0.018 $\textrm{d}^{-1}$, respectively, where $\Delta$$T$ is the time interval of observation
\citep{1978Ap&SS..56..285L,2019AJ....157...17L}. 

\begin{figure}
	\centering
	\includegraphics[width=0.5\textwidth, angle=0]{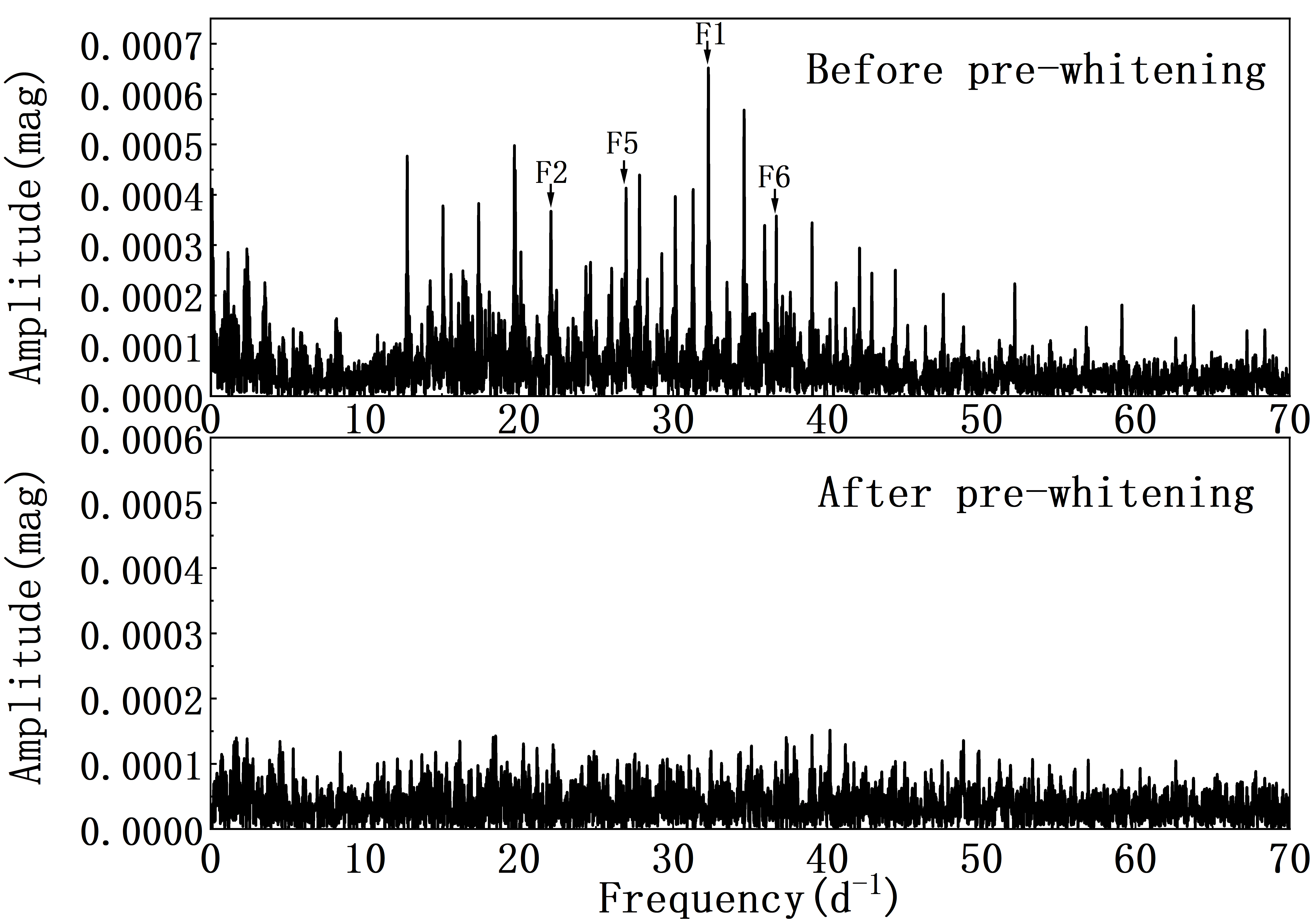}	
	\caption{The Fourier spectrum analysis displays the amplitude spectra both before and after pre-whitening using PERIOD04.}
	\label{fig:figure6}%
\end{figure}

Since frequencies above 70 $\textrm{d}^{-1}$ were not detected, we began with continuous pre-whitening of frequency peaks within the range of 0 to 70 $\textrm{d}^{-1}$. The amplitude spectra before and after pre-whitening are displayed in the upper and lower panels of Fig. \ref{fig:figure6}, respectively. If the spacing between two frequencies is smaller than the frequency resolution, we treat them as a single frequency and select the one with the higher signal-to-noise ratio (SNR). A frequency is considered a combination frequency if the amplitudes of the two parent frequencies exceed the amplitude of the anticipated combination frequency, and the discrepancy between the detected and projected frequencies is less than $\delta$$f$
\citep{2021MNRAS.505.6166S,2024NewA..11102234L}. A total of 26 frequencies with an SNR greater than 4.0 were extracted
\citep{1993A&A...271..482B} and were listed in Table \ref{tab:tab 4}, The errors in frequency, amplitude, and phase were calculated using the method of 
\cite{1999DSSN...13...28M}, The orbital frequency is F0 = 1.1590815 $\textrm{d}^{-1}$. The pulsating period of F1 (approximately 0.03 d) is near the result of 0.028 d given by 
\cite{2017MNRAS.465.1181L}.

\begin{table*}
\begin{center}
\footnotesize
\caption{The pulsating frequencies detected for CZ Aqr. \label{tab:tab 4}}
\setlength{\tabcolsep}{0.9cm}{
       \begin{tabular*}{\linewidth}{@{}cccccc@{}}
\hline
Label & Frequency($d^{-1}$) & Amplitude(mag) & Phase(rad/2$\pi$) & Combination & S/N  \\
\hline
F1    & 32.2825(11)    & 0.00050(2)     & 0.0991(78)    & F1          & 14   \\
F2    & 22.0744(12)    & 0.00047(2)     & 0.1319(83)    & 19F0   & 10.2 \\
F3    & 33.6106(12)    & 0.00045(2)     & 0.3909(86)    & 29F0        & 10.8 \\
F5    & 26.9465(13)    & 0.00042(2)     & 0.7672(93)    & F5          & 10   \\
F6    & 36.6912(14)    & 0.00040(2)     & 0.6039(99)    & F6          & 10.2 \\
F7    & 38.2514(15)    & 0.00037(2)     & 0.0567(107)   & 33F0        & 10.3 \\
F8    & 3.5128(16)     & 0.00033(2)     & 0.3619(117)   & 3F0         & 7.4  \\
F9    & 34.6023(17)    & 0.00033(2)     & 0.4388(119)   & 2F0+F1      & 6.2  \\
F10   & 40.5657(17)    & 0.00032(2)     & 0.8297(123)   & 35F0        & 8.4  \\
F11   & 42.0896(19)    & 0.00029(2)     & 0.1810(135)   & 68F0-F6     & 7.1  \\
F12   & 16.3617(19)    & 0.00029(2)     & 0.4253(135)   & 42F0-F1     & 6.2  \\
F13   & 20.1216(19)    & 0.00029(2)     & 0.4688(136)   & 49F0-F6     & 5.7  \\
F14   & 15.5923(20)    & 0.00028(2)     & 0.7922(139)   & 97F0-3F1    & 6.1  \\
F15   & 16.5590(21)    & 0.00027(2)     & 0.9660(145)   & 2F6-49F0    & 6.5  \\
F16   & 35.9928(21)    & 0.00026(2)     & 0.7020(152)   & 31F0         & 6.7  \\
F17   & 33.4833(23)    & 0.00024(2)     & 0.5305(164)   & F0+F1       & 5.5  \\
F18   & 26.0080(25)    & 0.00022(2)     & 0.5316(176)   & 4F1-89F0    & 6.2  \\
F19   & 25.8250(26)    & 0.00021(2)     & 0.1674(182)   & 2F6-41F0    & 6.6  \\
F20   & 26.8728(27)    & 0.00021(2)     & 0.3333(187)   & 3F16-70F0   & 4.7  \\
F21   & 26.6642(27)    & 0.00021(2)     & 0.9532(189)   & 23F0        & 4.9  \\
F22   & 15.0722(29)    & 0.00019(2)     & 0.6456(202)   & 13F0        & 4.1  \\
F23   & 16.9869(30)    & 0.00019(2)     & 0.4796(209)   & F6-17F0     & 5.2  \\
F24   & 5.8973(31)     & 0.00018(2)     & 0.6338(221)   & 5F0    & 5.1  \\
F25   & 17.5761(32)    & 0.00017(2)     & 0.5162(224)   & 3F16-78F0   & 4.3  \\
F26   & 21.2965(33)    & 0.00017(2)     & 0.0590(229)   & 4F1-93F0    & 4.9  \\
F27   & 32.1386(34)    & 0.00017(2)     & 0.5547(235)   & 2F1-28F0    & 4.2  \\
\hline
\end{tabular*}}
\begin{tablenotes} 
		\item F0 is the frequency of the orbital period.
     \end{tablenotes}
\end{center}
\end{table*} 

We calculated the pulsation constant $Q$ for the detected frequencies using equation (\ref{equation(14)}):
\begin{eqnarray}\label{equation(14)}
Q=P_{osc}\sqrt{{\rho}/{\rho_{\odot}}}, 
\end{eqnarray}
where $P_{osc}$ is the period of oscillation and $\rho$ = 0.27465 (±0.00043), $\rho_{\odot}$ is the mean density of the primary star deduced by the W-D program (see Table \ref{tab:tab 2}). The three possible independent frequencies were identified and matched with the theoretical model results for $M$ = 1.5 M$_{\odot}$
\citep{1981ApJ...249..218F}. The potential pulsation modes for each frequency were determined and were listed in Table \ref{tab:tab 5}.

\begin{figure}
	\centering
	\includegraphics[width=0.55\textwidth, angle=0]{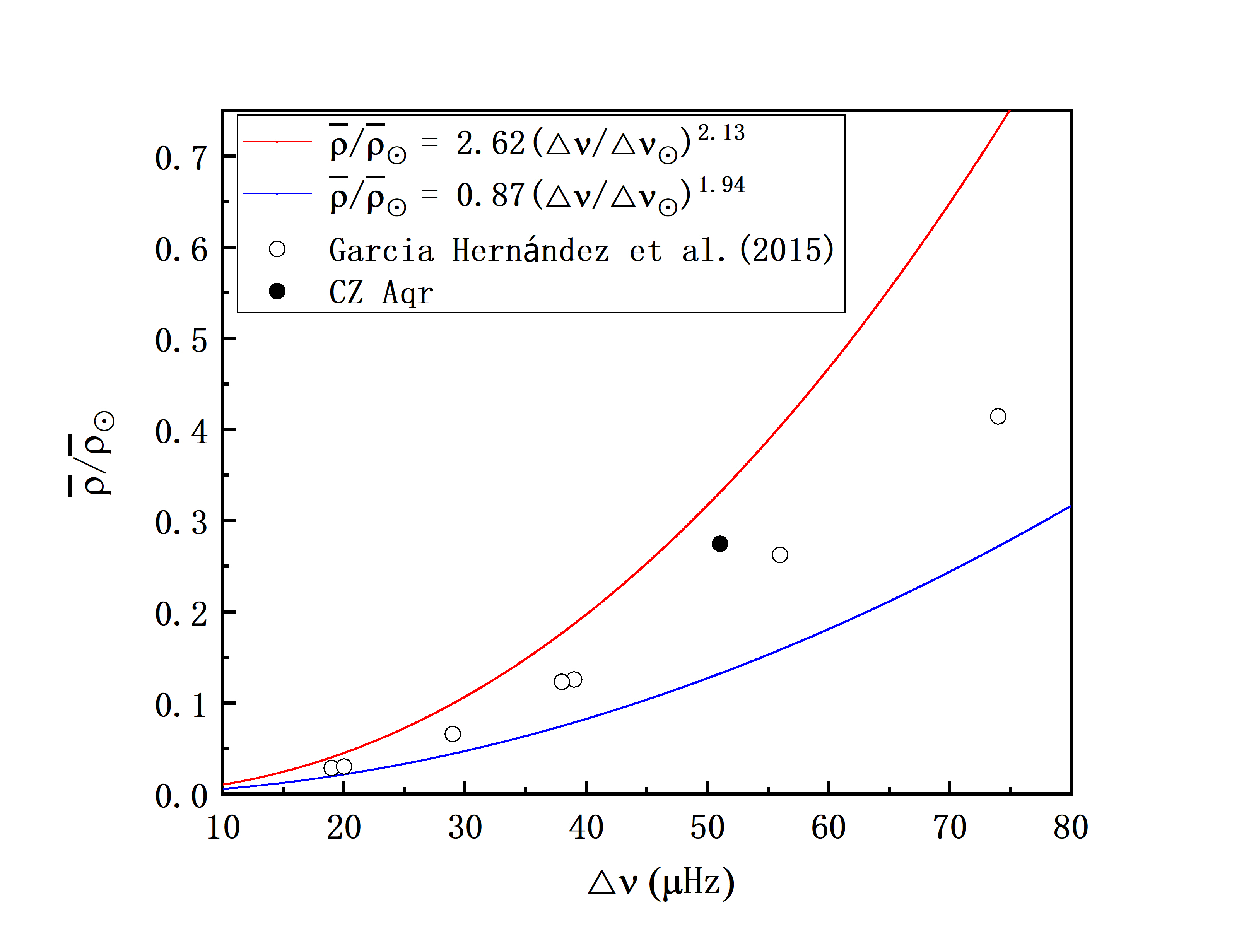}	
	\caption{The position of CZ Aqr in the large separation–mean density relation diagram for eclipsing binaries with a $\delta$ Sct component.}
	\label{fig:figure7}%
\end{figure}

The frequency difference between the two radial modes is $\Delta$$\nu$ = F6 - F1 = 4.4087\,($\pm$0.0020) $\textrm{d}^{-1}$ (51.0266 $\mu$Hz), which follows the basic equation
\citep{2015ApJ...811L..29G}:

\begin{eqnarray}\label{equation(15)}
\bar{\rho} / \bar{\rho}_{\odot}=1.55_{-0.68}^{+1.07}\left(\Delta \nu / \Delta \nu_{\odot}\right)^{2.035 \pm 0.095},
\end{eqnarray}
where $\Delta \nu_{\odot}$ = 134.8 $\mu$Hz is the large separation of the Sun. Its position in the large separation–mean density relation diagram is
shown in Fig. \ref{fig:figure7}, where the red and blue solid lines represent the boundary values of equation (13). The open circles represent the eclipsing binaries with a $\delta$ Sct component, which have been collected by 
\cite{2015ApJ...811L..29G}. This figure shows that CZ Aqr is consistent with the relation constructed by them, indicating that the two radial modes exist.

\begin{figure}
	\centering
	\includegraphics[width=0.55\textwidth, angle=0]{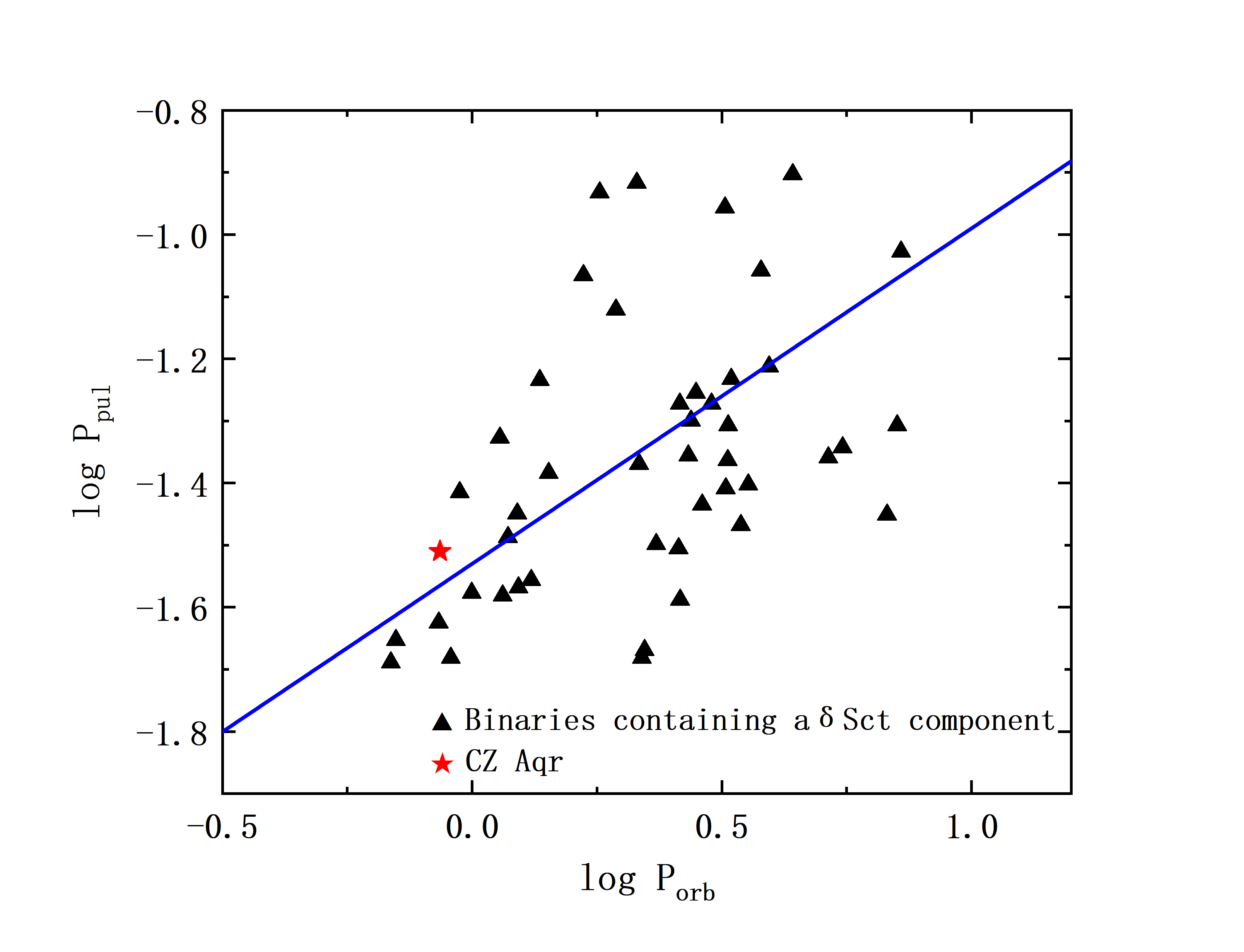}	
	\caption{The position of CZ Aqr in the $\log$ $P_{pul}$ versus $\log$ $P_{orb}$ diagram for binaries containing a $\delta$ Sct component with $P_{orb}$ $\textless$ 13 $d$, as presented by
 \cite{2017MNRAS.465.1181L}.}
	\label{fig:figure8}%
\end{figure}

The detected frequencies of F1 and F6 are found to follow the empirical relation between $P_{pul}$ and $P_{orb}$ for the semidetached binaries with a $\delta$ Sct-type star
\citep{2017MNRAS.465.1181L}:
\begin{eqnarray}\label{equation(16)}
\log P_{pul}=-1.53(3)+0.54(8) \log P_{orb}, r=0.62,
\end{eqnarray}
where $r$ is the correlation coefficient. The relationship between $P_{pul}$ and $P_{orb}$ for binaries containing a $\delta$ Sct component is displayed in Fig. \ref{fig:figure8}. 

\begin{table}
\caption{The independent pulsating frequencies and their corresponding pulsation modes. \label{tab:tab 5}}
\setlength{\tabcolsep}{2.0mm}{
       \begin{tabular*}{\linewidth}{@{}ccccc@{}}
\hline
Label & Frequency($d^{-1}$) & Q            & l-degree & Mode  \\
\hline
F1    & 32.2825(11)    & 0.016234(13) & 0        & R 3H  \\
F5    & 26.9465(13)    & 0.019449(15) &         &       \\
F6    & 36.6912(14)    & 0.014283(11) & 0        & R 4H  \\
\hline
\end{tabular*}}
\end{table}

\section{Discussions and Conclusions}
\subsection{Analysis of Variations in The Orbital Period}

In this paper, a total of 290 light minima spanning 108 years were used for the analysis of changes in the orbital period. We found that the long-term period decreases with $\beta$ = -3.09 (±0.63)$\times$$10^{-8}$ d $\textrm{yr}^{-1}$. Given the configuration of CZ Aqr, this may be attributed to angular momentum loss caused by magnetic stellar wind. When material escapes from the star’s surface in the form of stellar wind, angular momentum is carried away and lost. To estimate the total mass loss of the system, we used the Alfvén-driven mass transfer equation
\citep{1991MNRAS.253....9T}:

\begin{eqnarray}\label{equation(6)}
\frac{ \dot{P}}{P} = -\frac{2 \dot{M}}{M} + \frac{3 \dot{M}_{2}\left(M_{2}-M_{1}\right)}{M_{1} M_{2}} + 2\left(\frac{R_{A}}{d}\right)^{2} \frac{M \dot{M}}{M_{1} M_{2}}
\end{eqnarray}

where $M_{1}$ and $M_{2}$ are the masses of the primary and secondary components, respectively. Equation (\ref{equation(6)}) consists of three parts: the first part represents the change in the orbital period due to the total mass loss of the system, the second part represents the effect of mass exchange on the variation of the orbital period, and the third part represents the change in the orbital period due to the additional angular momentum lost as a result of magnetic field coupling. This coupling ensures that the stellar wind material remains in synchronous rotation with the star up to a distance from the star's surface, known as the Alfvén radius $R_{A}$, which can be up to ten times the radius of the star. Combining the $O-C$ analysis results with the absolute parameters obtained in the previous section, we estimated the total mass loss rate as approximately $\dot{M}$ = 4.54$\times$$10^{-9}$ M$_{\odot}$ $\textrm{yr}^{-1}$. 
\cite{2014ASPC..482..127D} also stated that the hot spot model is the best way to explain mass loss from such Algol-type systems. As for the cyclical variation, it could be explained by the LTTE due to the presence of a third body. Therefore, we use the following equation and the estimated absolute parameters in Section 3 to estimate the mass of a potential third body:

\begin{eqnarray}\label{equation(7)}
f(\mathrm{m})=\frac{4 \pi^{2}}{G P_{3}^{2}} \times\left(a_{12} \sin i_{3}\right)^{3}=\frac{\left(M_{3} \sin i_{3}\right)^{3}}{\left(M_{1}+M_{2}+M_{3}\right)^{2}}
\end{eqnarray}

where $a_{12} \sin i_{3}$ = $A$ $\times$ $c$ (see Table \ref{tab:tab 1}). The mass function $f$(m) = 0.0042 (±0.0012) M$_{\odot}$, and the minimum mass of the third body $M_{3}$ = 0.312 (±0.021) M$_{\odot}$. All of these physical parameters of the third body are listed in Table \ref{tab:tab 1}. Since the third body has a negligible light contribution that cannot be detected, the W-D program has no convergent solutions for the third light, which is probably a cool dwarf star. The presence of a third body in close binaries is very common
\citep{2006AJ....131.2986P,2021MNRAS.508.6111L}.

\begin{figure}
    \centering
    \includegraphics[width=0.5\textwidth, angle=0]{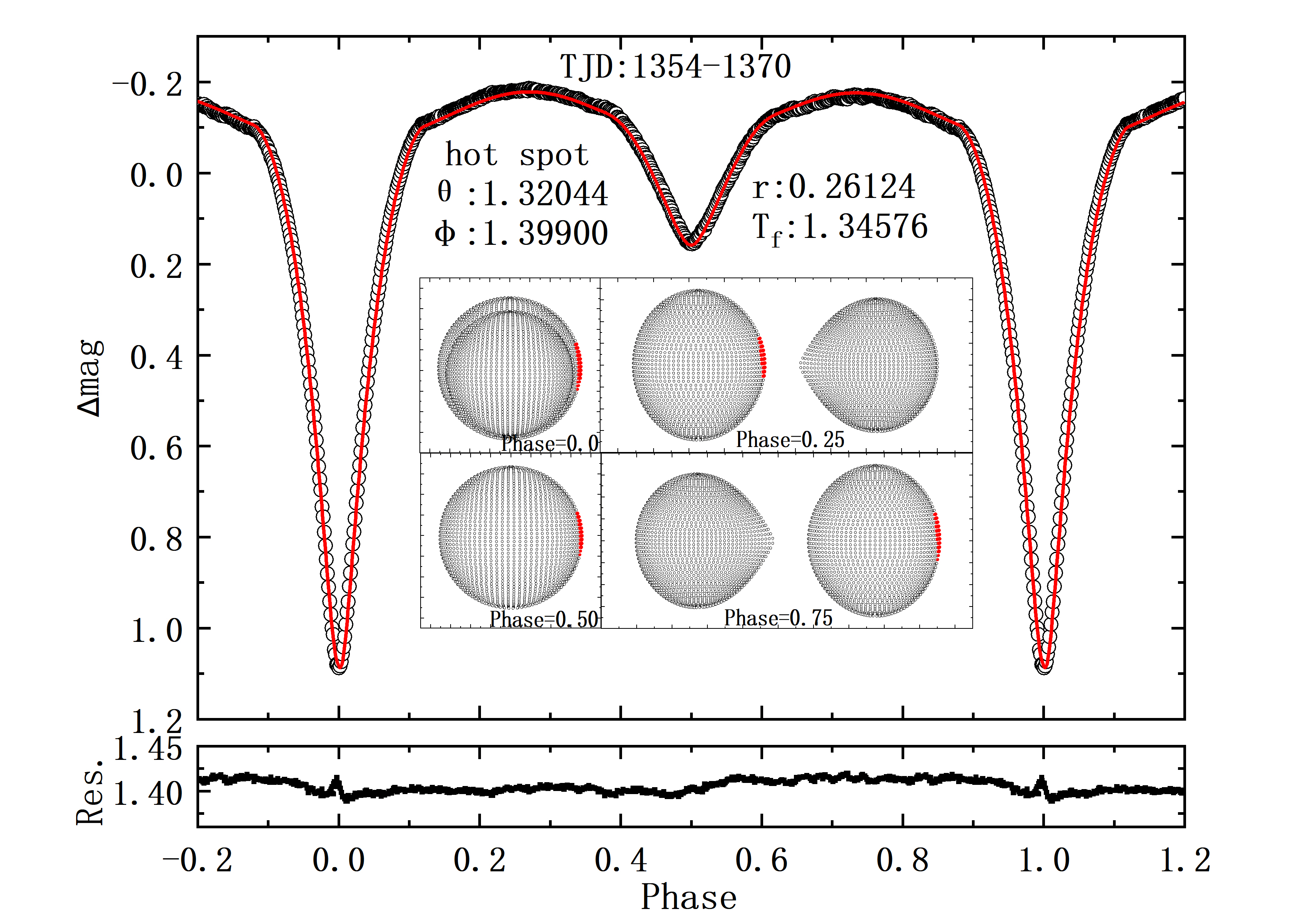}
    \hspace{0.001in}
   \includegraphics[width=0.5\textwidth, angle=0]{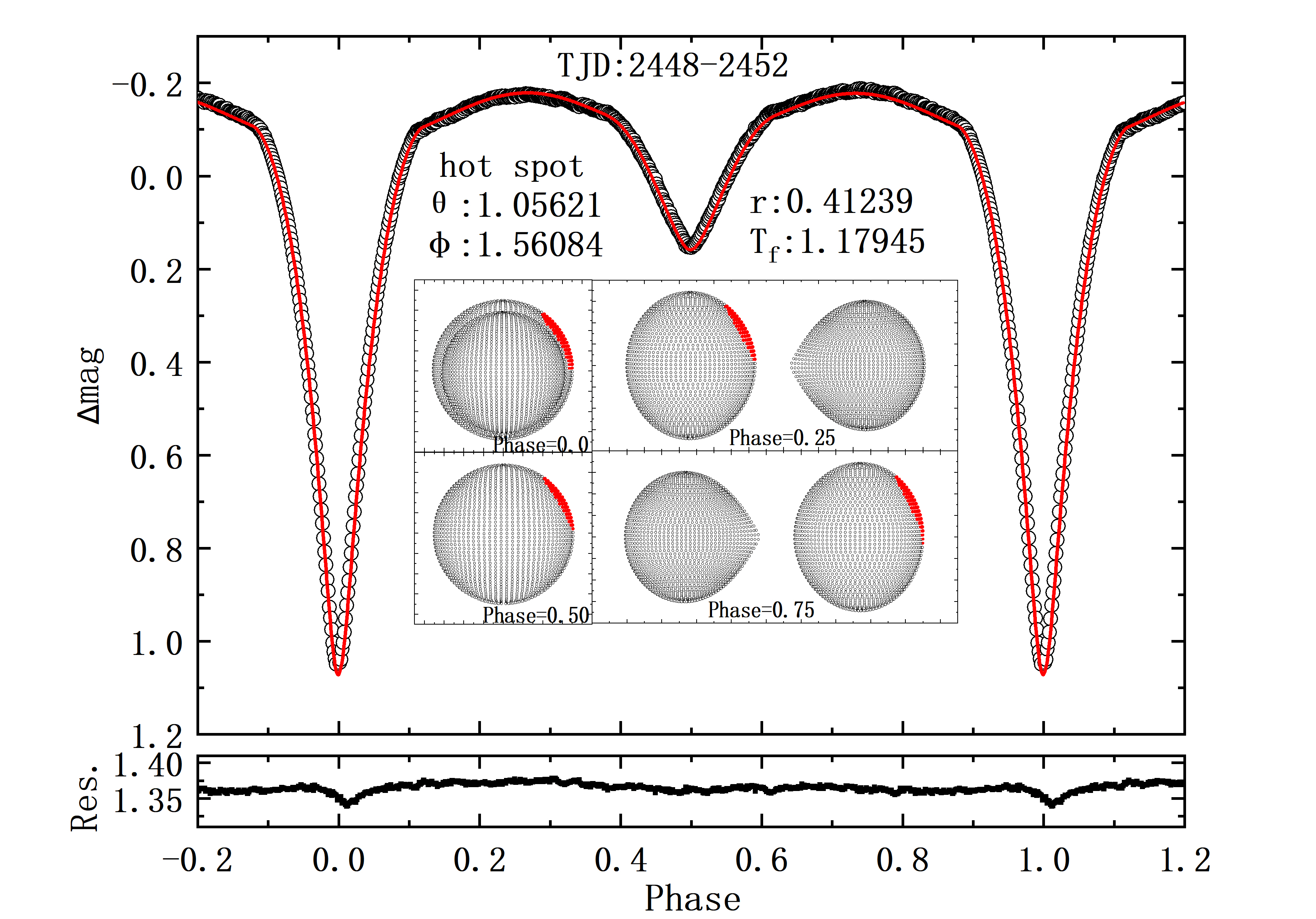}  
    \caption{The hot spot solutions for CZ Aqr, based on the basic solution, are shown with red dots representing the evolving hot spot on the primary component. The spot parameters $\theta$, $\phi$, $r$, and $T_{f}$ represent latitude (in radian), longitude (in radian), radius (in radian), and temperature factor, respectively. They are detailed in the figure. The solutions for the positive and negative O'Connell effects are displayed in the upper and lower panels, respectively. The residuals are displayed in the bottom panel.}
    \label{fig:figure9}%
\end{figure}

\subsection{Summary}
The O'Connell effect
\citep{1969CoKon..65..457M,2022yCat..22620010K} indicates asymmetry in the light curve, where the light maximum at phase 0.25 (MAX I) differs from the light maximum at phase 0.75 (MAX II). The positive and negative O’Connell effects are displayed in the upper and lower panels of Fig. \ref{fig:figure9}, respectively. The case of upper panel represents a better spot model for CZ Aqr because it has a smaller $\chi^{2}$ value of 0.0017. The spot parameters $\theta$, $\phi$, $r$, and $T_{f}$ represent latitude (in radian), longitude (in radian), radius (in radian), and temperature factor, respectively. This effect is reversed due to the interactions within the binary system. The appearance of a hot spot on the surface of the primary star may be attributed to the transfer of mass from the secondary component to the primary one.
\begin{figure}
    \centering
   \includegraphics[width=0.5\textwidth, angle=0]{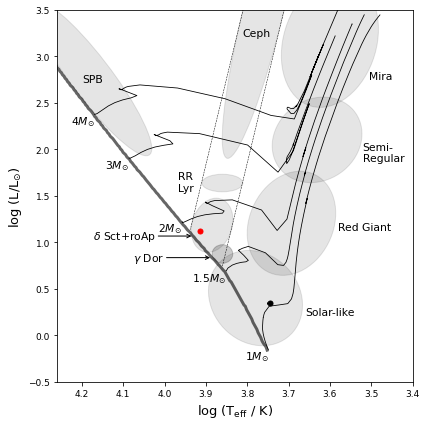}  
    \caption{The positions of the primary component (red dot) and the secondary component (black dot) in the H-R diagram
    \citep{2010aste.book.....A}.}
    \label{fig:figure10}%
\end{figure}

In the Mass-Radius (M-R) relationship, A-type close binaries typically show that the primaries are evolving towards the terminal-age main sequence (TAMS), while the secondaries pass through the ZAMS
\citep{2016ApJ...823..102C,2003ASPC..292..129N}. However, most binary members with $\delta$ Sct stars commonly lie within the MS band and approach the ZAMS boundary, with the secondaries gradually moving away from the TAMS
\citep{2012MNRAS.422.1250L}. This behavior may be due to the interactions between the two components.

Regarding the interaction between these two components, we discover that the secondary component nearly fills its Roche lobe and transfers mass to the primary component. The secondary, having undergone more evolution, suggests that it was once the more massive primary star. This star evolved more quickly and transferred mass to the current primary star
\citep{1955ApJ...121...71C}. This behavior aligns with the Algol paradox, which posits that the more massive star expands its outer layers as it evolves, eventually filling its Roche lobe and transferring mass to its less massive companion, thereby reversing the mass ratio
\citep{1998A&AT...15..357P}. The position of the two components are shown in Fig. \ref{fig:figure10}. The primary star (red dot) lies within the pulsating instability region of $\delta$ Sct-type star, while the secondary component (black dot) is located in the subgiant evolutionary stage.

In conclusion, CZ Aqr is an evolving Algol-type binary system. The W-D program results indicate that the secondary component nearly fills its Roche lobe and is transferring mass rapidly to the primary component. The long-term decrease in the orbital period observed in the $O-C$ analysis may be due to angular momentum loss driven by magnetic stellar wind. We estimated the corresponding total mass loss, and the mass loss from Algol-type systems can also be explained by the hot spot 
\citep{2014ASPC..482..127D}. The cyclical variations may be attributed to the presence of a third body and possibly a fourth body, with these bodies potentially following a 2:7 resonance orbit around the CZ Aqr binary pair. The primary component of the binary system is identified as a $\delta$ Sct-type star, which may exhibit radial mode. Compared to previous studies, our results reveal that the eclipsing Algol-type system comprises a $\delta$ Sct primary star and a subgiant star within a quadruple system. We have estimated the total mass loss rate, identified another potential celestial body and a possible resonant orbit, pinpointed the positions of the binary components in the H-R diagram, and identified two potential radial modes. Consequently, CZ Aqr presents a valuable opportunity for studying the formation and evolution of binaries with a $\delta$ Sct-type star.

\section*{Acknowledgments}
This work is supported by the International Cooperation Projects of the National Key R$\&$D Program (No. 2022YFE0127300), the National Natural Science Foundation of China (No. 11933008), 2022 CAS ``Light of West China" Program, the Young Talent Project of  ``Yunnan Revitalization Talent Support Program" in Yunnan Province, the basic research project of Yunnan Province (Grant No. 202201AT070092).
The TESS data presented in this paper were obtained from the Mikulski Archive for Space Telescopes (MAST) at the Space Telescope Science Institute (STScI). The specific observations analyzed can be accessed via\dataset[http://dx.doi.org/10.17909/c63n-np75].. STScI is operated by the Association of Universities for Research in Astronomy, Inc. Support to MAST for these data is provided by the NASA Office of Space Science. Funding for the TESS mission is provided by the NASA Explorer Program. This work has made use of data from the European Space Agency (ESA) mission Gaia. Processed by the Gaia Data Processing and Analysis Consortium. Funding for the DPAC has been provided by national institutions, in particular, the institutions participating in the Gaia Multilateral Agreement.

%





\bibliography{sample631}{}
\bibliographystyle{aasjournal}



\begin{table*}
\begin{center}
\begin{threeparttable}
\footnotesize
\caption{The light minima of CZ Aqr are presented.\label{tab:tab 6}}
\setlength{\tabcolsep}{0.5mm}{
       \begin{tabular*}{\linewidth}{@{}llllllllllll@{}}
\hline
\multicolumn{1}{c}{HJD}           & Ref. & \multicolumn{1}{c}{HJD}           & Ref. & \multicolumn{1}{c}{HJD}                & Ref. & \multicolumn{1}{c}{HJD}                & Ref. & \multicolumn{1}{c}{HJD}                & Ref. & \multicolumn{1}{c}{HJD}               & Ref. \\
\hline 
2420773.36000 & (3)  & 2442398.28000 & (5)  & 2445904.50800      & (5)  & 2452952.32480(30)  & (8)  & 2458373.83271(08)  & (1)  & 2459851.71890     & (5)  \\
2420786.34000 & (3)  & 2442404.32100 & (5)  & 2445904.51400      & (5)  & 2452963.54030      & (5)  & 2458374.69538(11)  & (1)  & 2460183.01349(11) & (1)  \\
2425506.46000 & (3)  & 2442417.25800 & (5)  & 2445916.58800      & (5)  & 2453227.54700      & (5)  & 2458375.55826(08)  & (1)  & 2460183.87630(10) & (1)  \\
2425512.45000 & (3)  & 2442417.25900 & (5)  & 2445945.05600      & (3)  & 2453238.75680      & (5)  & 2458376.42109(11)  & (1)  & 2460184.73891(14) & (1)  \\
2425864.46000 & (3)  & 2442712.32400 & (5)  & 2445987.33300      & (5)  & 2453314.67910      & (5)  & 2458377.28364(09)  & (1)  & 2460185.60177(11) & (1)  \\
2425883.47000 & (3)  & 2442782.21200 & (5)  & 2446262.54000      & (5)  & 2453426.83663(115) & (7)  & 2458378.14642(13)  & (1)  & 2460186.46450(14) & (1)  \\
2425909.33000 & (3)  & 2442962.52800 & (5)  & 2446268.58400      & (5)  & 2453592.49300      & (5)  & 2458379.00914(11)  & (1)  & 2460187.32710(13) & (1)  \\
2426242.37000 & (3)  & 2442987.55600 & (5)  & 2446319.49000      & (5)  & 2453634.75950      & (5)  & 2458379.87200(08)  & (1)  & 2460189.05256(09) & (1)  \\
2426267.36000 & (3)  & 2442993.56700 & (5)  & 2446377.29400      & (5)  & 2453653.74200      & (5)  & 2458380.73465(12)  & (1)  & 2460189.91537(18) & (1)  \\
2426625.38000 & (3)  & 2443371.47200 & (3)  & 2446428.19700      & (5)  & 2453673.58300      & (5)  & 2459058.85450      & (5)  & 2460190.77815(10) & (1)  \\
2426651.30000 & (3)  & 2443371.47200 & (5)  & 2446627.49500      & (5)  & 2454035.94300      & (2)  & 2459089.05087(104) & (1)  & 2460191.64078(09) & (1)  \\
2427413.97200 & (3)  & 2443735.55600 & (5)  & 2446646.48000      & (5)  & 2454075.62300      & (5)  & 2459089.91340(77)  & (1)  & 2460192.50382(17) & (1)  \\
2430003.11700 & (3)  & 2443806.29200 & (5)  & 2446742.23900      & (5)  & 2454094.60330      & (5)  & 2459090.77593(78)  & (1)  & 2460195.95458(12) & (1)  \\
2430969.37000 & (3)  & 2443806.29700 & (5)  & 2447010.55300      & (5)  & 2454161.89728(84)  & (7)  & 2459091.63863(64)  & (1)  & 2460196.81734(12) & (1)  \\
2430976.26900 & (3)  & 2443806.30100 & (5)  & 2447029.53400      & (5)  & 2454317.19160      & (5)  & 2459092.50180(148) & (1)  & 2460197.68020(07) & (1)  \\
2431001.28800 & (3)  & 2443831.31400 & (5)  & 2447157.22200      & (5)  & 2454452.64270      & (5)  & 2459093.36426(68)  & (1)  & 2460198.54291(13) & (1)  \\
2432823.42200 & (3)  & 2443837.35300 & (5)  & 2447471.27400      & (5)  & 2454702.83900      & (5)  & 2459094.22668(72)  & (1)  & 2460199.40546(13) & (1)  \\
2433514.48700 & (3)  & 2443838.22400 & (5)  & 2447739.60100      & (5)  & 2454710.60306(45)  & (7)  & 2459095.09030(104) & (1)  & 2460200.26822(10) & (1)  \\
2433539.51300 & (3)  & 2443863.23500 & (5)  & 2447822.40000      & (5)  & 2455057.42720(91)  & (7)  & 2459095.95282(94)  & (1)  & 2460201.13115(12) & (1)  \\
2433872.52800 & (3)  & 2443888.25800 & (5)  & 2448098.49100      & (5)  & 2455504.33180(20)  & (9)  & 2459096.81523(85)  & (1)  & 2460201.99393(12) & (1)  \\
2434576.54100 & (3)  & 2444087.55500 & (5)  & 2448174.40500      & (5)  & 2455505.19320(30)  & (9)  & 2459097.67792(58)  & (1)  & 2460202.85661(15) & (1)  \\
2435011.36600 & (3)  & 2444164.33900 & (5)  & 2448488.44600      & (5)  & 2455508.64570(20)  & (10) & 2459098.54076(89)  & (1)  & 2460203.71936(09) & (1)  \\
2435401.33300 & (3)  & 2444208.33000 & (5)  & 2448564.37000      & (5)  & 2455536.25320(10)  & (9)  & 2459102.85529(91)  & (1)  & 2460204.58225(10) & (1)  \\
2435721.41600 & (3)  & 2444458.53600 & (5)  & 2448564.37400      & (5)  & 2456512.88300(10)  & (5)  & 2459103.71764(69)  & (1)  & 2460205.44486(11) & (1)  \\
2435778.35700 & (3)  & 2444468.88520 & (4)  & 2448859.43300      & (5)  & 2457235.86670(10)  & (5)  & 2459104.57990(63)  & (1)  & 2460209.75859(10) & (1)  \\
2440198.24600 & (5)  & 2444484.41900 & (5)  & 2449216.61200      & (5)  & 2458067.55710      & (5)  & 2459105.44256(76)  & (1)  & 2460210.62135(07) & (1)  \\
2440537.29900 & (5)  & 2444489.59600 & (5)  & 2449600.53700      & (5)  & 2458111.55780      & (5)  & 2459106.30589(72)  & (1)  & 2460211.48413(10) & (1)  \\
2440799.58400 & (5)  & 2444497.35600 & (5)  & 2449633.32500      & (5)  & 2458354.85223(08)  & (1)  & 2459107.16879(108) & (1)  & 2460212.34692(11) & (1)  \\
2440921.25300 & (5)  & 2444528.42200 & (5)  & 2449984.46400      & (5)  & 2458355.71497(09)  & (1)  & 2459108.03010(97)  & (1)  & 2460213.20962(08) & (1)  \\
2441241.33300 & (5)  & 2444566.38000 & (5)  & 2450291.60200      & (5)  & 2458356.57774(09)  & (1)  & 2459108.89427(106) & (1)  & 2460214.07240(10) & (1)  \\
2441522.58900 & (5)  & 2444586.22900 & (5)  & 2450336.47000      & (5)  & 2458357.44051(07)  & (1)  & 2459109.75705(59)  & (1)  & 2460214.93508(09) & (1)  \\
2441535.53700 & (5)  & 2444598.30200 & (5)  & 2450425.33300      & (5)  & 2458358.30322(10)  & (1)  & 2459110.61942(95)  & (1)  & 2460215.79787(10) & (1)  \\
2441605.41300 & (5)  & 2444810.53200 & (5)  & 2450700.55700      & (5)  & 2458359.16610(09)  & (1)  & 2459111.48208(71)  & (1)  & 2460216.66073(10) & (1)  \\
2441631.29300 & (5)  & 2444810.53300 & (5)  & 2450989.58000      & (5)  & 2458360.02870(11)  & (1)  & 2459447.95304(11)  & (1)  & 2460222.69993(07) & (1)  \\
2441637.32600 & (5)  & 2444810.53400 & (5)  & 2451123.30400      & (5)  & 2458360.89160(13)  & (1)  & 2459448.81554(10)  & (1)  & 2460223.56262(09) & (1)  \\
2441650.27600 & (5)  & 2444925.28000 & (5)  & 2451436.47600      & (5)  & 2458361.75427(11)  & (1)  & 2459449.67834(08)  & (1)  & 2460224.42518(07) & (1)  \\
2441874.58500 & (5)  & 2445193.60600 & (5)  & 2451454.59800      & (5)  & 2458361.75460      & (5)  & 2459450.54110(09)  & (1)  & 2460225.28811(06) & (1)  \\
2441900.46800 & (5)  & 2445194.46400 & (5)  & 2451756.54900      & (5)  & 2458362.61704(08)  & (1)  & 2459451.40377(10)  & (1)  & 2460226.15080(08) & (1)  \\
2441976.39600 & (5)  & 2445232.42800 & (5)  & 2451839.38600      & (5)  & 2458363.47991(11)  & (1)  & 2459461.75687(08)  & (1)  & 2460227.01358(09) & (1)  \\
2441989.33200 & (5)  & 2445238.45800 & (5)  & 2451849.73500      & (5)  & 2458364.34250(09)  & (1)  & 2459462.61944(08)  & (1)  & 2460227.87643(10) & (1)  \\
2442008.31700 & (5)  & 2445251.39700 & (5)  & 2452065.41858(178) & (7)  & 2458365.20519(09)  & (1)  & 2459463.48221(08)  & (1)  & 2460228.73902(07) & (1)  \\
2442258.51600 & (5)  & 2445251.40100 & (5)  & 2452165.50100      & (5)  & 2458366.06804(10)  & (1)  & 2459464.34491(08)  & (1)  & 2460229.60178(07) & (1)  \\
2442289.57300 & (5)  & 2445276.42000 & (5)  & 2452189.65570      & (5)  & 2458366.93069(08)  & (1)  & 2459465.20777(08)  & (1)  & 2460230.46451(10) & (1)  \\
2442289.57500 & (5)  & 2445277.28400 & (5)  & 2452229.34300      & (5)  & 2458368.65629(12)  & (1)  & 2459466.07044(08)  & (1)  & 2460231.32734(09) & (1)  \\
2442296.48100 & (5)  & 2445328.21000 & (5)  & 2452460.56600      & (5)  & 2458369.51912(12)  & (1)  & 2459466.93338(11)  & (1)  & 2460232.18988(09) & (1)  \\
2442303.37900 & (5)  & 2445622.38100 & (5)  & 2452550.28425(60)  & (7)  & 2458370.38171(08)  & (1)  & 2459467.79600(09)  & (1)  &                   &      \\
2442303.37900 & (5)  & 2445635.32900 & (5)  & 2452850.52400(400) & (6)  & 2458371.24460(10)  & (1)  & 2459468.65877(07)  & (1)  &                   &      \\
2442367.22500 & (5)  & 2445635.33100 & (5)  & 2452887.61885(50)  & (7)  & 2458372.10710(07)  & (1)  & 2459469.52134(09)  & (1)  &                   &      \\
2442385.34700 & (5)  & 2445705.21800 & (5)  & 2452931.61900      & (5)  & 2458372.96995(08)  & (1)  & 2459470.38428(11)  & (1)  &                   &    \\  
\hline
\end{tabular*}}

 \begin{tablenotes} 
		\item Reference. (1) This paper; (2) GCVS; (3) \cite{1986IBVS.2865....1B}; (4) 
  \cite{1982IBVS.2185....1W}; (5) $O-C$ gateway; (6)
  \cite{2003IBVS.5438....1D}; (7) ASAS; (8) 
  \cite{2006IBVS.5676....1K}; (9) 
  \cite{2010IBVS.5958....1L}; (10) \cite{2011IBVS.5960....1D}.
     \end{tablenotes} 
\end{threeparttable} 
\end{center}
\end{table*}

\end{document}